\documentclass[12pt]{iopart}
\usepackage{amssymb}
\begin{document}
\title{Can Brans-Dicke
scalar field account for dark energy and dark matter?}
\author{M. Ar\i k and M. C. \c{C}al\i k}
\begin{abstract}
By using a linearized non-vacuum late time solution in Brans-Dicke cosmology
we account for the seventy five percent dark energy contribution but not for
approximately twenty-three percent dark matter contribution to the present
day energy density of the universe.
\end{abstract}
\address{Bo\~{g}azi\c{c}i Univ., Dept. of Physics, Bebek, Istanbul, Turkey} %
\eads{\mailto{arikm@boun.edu.tr}, \mailto{cem.calik@boun.edu.tr}}

Our universe seems, according to the present-day evidence, to be
spatially flat and to possess a non vanishing cosmological constant
\cite{correl, Liddle}. For a flat matter dominated universe,
cosmological measurements \cite{knop} imply that the fraction
$\Omega _{\Lambda }$ of the contribution of the cosmological
constant $\Lambda $ to present energy density of the universe is
$\Omega _{\Lambda }\sim $ $0.75$. In standard cosmology $\Omega
_{\Lambda }$ would be induced by a cosmological constant which is a
dimensionful parameter with units of (length)$^{-2}$. From the point
of view of classical general relativity, there is no preferred
choice for what the length scale defined by $\Lambda $ might be.
Particle physics, however, gives a different point of view to the
issue. The cosmological constant turns out to be a measure of the
energy density of the vacuum and although we can not calculate the
vacuum energy with any confidence, this allows us to consider the
scales of various contributions to the cosmological constant. The
energy scale of the constituent(s) of $\Lambda $ which in Planck
units is approximated to $10^{-123}$ is problematic since it is
lower than the normal energy scale predicted by most particle
physics models. To solve this problem, a dynamical $\Lambda $
\cite{3} in the form of scalar field with some self interacting
potential \cite{4} can be considered and its slowly varying energy
density induces a cosmological constant. This idea called
\textquotedblleft \textit{quintessence\textquotedblright } \cite{3}
is similar to the inflationary phase of the early universe with the
difference that it evolves at a much lower energy density scale. The
energy density of this field has to evolve in such a way that it
becomes comparable with the mass density fraction $\Omega _{M}$ now.
This type of specific evolution, better known as \textquotedblleft
\textit{cosmic coincidence} \textquotedblright\ \cite{6} problem,
needs several constraints and fine tuning of parameters for the
potential used to model quintessence with minimally coupled scalar
field. To solve the cosmic coincidence problem, a new form of
quintessence field called the \textquotedblleft tracker
field\textquotedblright\ \cite{6} has been proposed. Such kind of
quintessence field is mainly based on an equation of motion with a
solution for such that for a wide range of \ initial conditions the
equation of motion converge to the same solution. This type of
solution is also called an 'attractor like' solution. There are a
number of quintessence models proposed. Most of these involve
minimally coupled scalar field with different potentials dominating
over the kinetic energy of the field. Purely exponential
\cite{7}-\cite{7.2} and inverse power law \cite{4}-\cite{6}
potentials have been extensively studied for quintessence fields to
solve the cosmic coincidence problem. However the fact that the
energy density is not enough to make up for the missing part of the
cosmological constant or that the $p/\rho \equiv \gamma $ value
found for the equation of state of quintessence is not in good
agreement with the observed results makes such an explanation
unlikely. The investigation of alternative models in which the
equation of state parameter $\gamma $ of the cosmological constant
evolves with time has been proposed due to the conceptual
difficulties associated with a cosmological constant
\cite{padmn}-\cite{gon}.

There have been quite a few attempts for treating this problem with
non-minimal coupled scalar fields. Studies made by Bartolo \textit{et al}
\cite{8}, Bertolami \textit{et al }\cite{9}\textit{, }Ritis \textit{et al}
\cite{10} have found tracking solutions in scalar tensor theories with
different types of power law potential. In another work, Sen \textit{et al}
\cite{11} have found the potential relevant to power law expansion in
Brans-Dicke (BD) cosmology and Ar\i k \textit{et al} \cite{12} have shown
that (BD) theory of gravity with the standard mass term potential $\frac{1}{2%
}m^{2}\phi ^{2}$ is a natural model to explain the rapid primordial
inflation and the observed slow late-time inflation.

In this paper we show that a linearized non-vacuum solution about the stable
cosmological vacuum solution with flat space-like section is capable of
explaining how the Hubble parameter evolves with the scale size of the
universe $a(t)$. In this framework, we also show that the standard Friedmann
equation changes into a form in which the power of the scale size term with $%
\Omega _{M}$ is corrected by an amount $1/\omega $%
\begin{equation}
\left( \frac{H}{H_{0}}\right) ^{2}=\Omega _{\Lambda }+\Omega _{M}\left(
\frac{a_{0}}{a}\right) ^{^{3+\frac{1}{\omega }}}  \label{hubble with w}
\end{equation}
where $\omega $ is the Brans-Dicke parameter with $\omega \gg 1$
\cite{omega, omega1}. Subsequently, under such a linearized
solution, we point out that only a very small part of the dark
matter can be accommodated into the contribution of the Brans-Dicke
scalar field.

In the context of (BD) theory \cite{13} with self interacting potential and
matter field, the action in the canonical form is given by%
\begin{equation}
S=\int d^{4}x\,\sqrt{g}\,\left[ -\frac{1}{8\omega }\,\phi ^{2}\,R+\frac{1}{2}%
\,g^{\mu \upsilon }\,\partial _{\mu }\phi \,\partial _{\nu }\phi -\frac{1}{2}%
m^{2}\phi ^{2}+L_{M}\right] .  \label{action*}
\end{equation}%
In particular we may expect that $\phi $ is spatially uniform, but varies
slowly with time. The nonminimal coupling term $\,\phi ^{2}\,R$ where $R$ is
the Ricci scalar, replaces with the Einstein-Hilbert term $\frac{1}{G_{N}}R$
in such a way that $G_{eff}^{-1}=\frac{2\pi }{\omega }\phi ^{2}$ where $%
G_{eff}$ is the effective gravitational constant as long as the dynamical
scalar field $\phi $ varies slowly. In units where $c=\hbar =1$, we define
Planck-length, $L_{p}$, in such a way that $L_{P}^{2}\phi _{0}^{2}=\omega
/2\pi $ where $\phi _{0}$ is the present value of the scalar field $\phi $.
Thus, the dimension of the scalar field is chosen to be $L_{P}^{-1}$ so that
$G_{eff}$ has a dimension $L_{P}^{2}$. The signs of the non-minimal coupling
term and the kinetic energy term are properly adopted to $(+---)$ metric
signature. The Lagrangian of the scalar field, in addition to non-minimal
coupling term and the kinetic term, is composed of a potential which
consists of only a standard mass term. $L_{M}$, on the other hand, is the
matter part of the Lagrangian which in accordance with the weak equivalence
principle is decoupled from $\phi $ as has been assumed in the original (BD)
theory. Excluding $\phi $, as the matter field, we consider a classical
perfect fluid with the energy-momentum tensor $T_{\nu }^{\mu }=diag\left(
\rho ,-p,-p,-p\right) $ where $\rho $ is the energy density, $p$ is the
pressure.

The gravitational field equations derived from the variation of the action (%
\ref{action*}) with respect to Robertson- Walker metric is
\begin{equation}
\frac{3}{4\omega }\,\phi ^{2}\,\left( \frac{\dot{a}^{2}}{a^{2}}+\frac{k}{%
a^{2}}\right) -\frac{1}{2}\,\dot{\phi}^{2}-\frac{1}{2}\,m^{2}\,\phi ^{2}+%
\frac{3}{2\omega }\,\frac{\dot{a}}{a}\,\dot{\phi}\,\phi =\rho _{M}
\label{des}
\end{equation}%
\begin{equation}
\frac{-1}{4\omega }\phi ^{2}\left( 2\frac{\ddot{a}}{a}+\frac{\dot{a}^{2}}{%
a^{2}}+\frac{k}{a^{2}}\right) -\frac{1}{\omega }\,\frac{\dot{a}}{a}\,\dot{%
\phi}\,\phi -\frac{1}{2\omega }\,\ddot{\phi}\,\phi -\left( \frac{1}{2}+\frac{%
1}{2\omega }\right) \,\dot{\phi}^{2}+\frac{1}{2}\,m^{2}\,\phi ^{2}=p_{M}
\label{pres}
\end{equation}%
\begin{equation}
\ddot{\phi}+3\,\frac{\dot{a}}{a}\,\dot{\phi}+\left[ m^{2}-\frac{3}{2\omega }%
\left( \frac{\ddot{a}}{a}+\frac{\dot{a}^{2}}{a^{2}}+\frac{k}{a^{2}}\right) %
\right] \,\phi =0  \label{fi}
\end{equation}%
where $k$\ is the curvature parameter with $k=-1$, $0$, $1$\ corresponding
to open, flat, closed universes respectively and $a\left( t\right) $ is the
scale factor of the universe (dot denotes $\frac{d}{dt}$). Since in the
standard theory of gravitation, the total energy density $\rho $ is assumed
to be composed of $\rho =\rho _{\Lambda }+\rho _{M}$ where $\rho _{\Lambda }$
is the energy density of the universe due to the cosmological constant which
in modern terminology is called as \textquotedblleft \textit{dark
energy\textquotedblright ,} the right hand sides of (\ref{des}, \ref{pres})
are adopted to the matter energy density term $\rho _{M}$ instead of $\rho $
and $p_{M}$ instead of $p$ where $M$ denotes everything except the $\phi $
field. The main reason behind doing such an organization is that whether if
the $\phi $ terms on the left-hand side of (\ref{des}) can accommodate a
contribution to due to what is called dark matter. In addition, the right
hand side of the $\phi $ equation (\ref{fi}) is set to be zero according to
the assumption imposed on the matter Lagrangian $L_{M}$ being independent of
the scalar field $\phi $. By defining the fractional rate of change of $\phi
$ as $F\left( a\right) =\dot{\phi}/\phi $ and the Hubble parameter as $%
H\left( a\right) =\dot{a}/a$, we rewrite the left hand-side of the field
equations (\ref{des}-\ref{fi}) in terms of $H(a)$, $F(a)$ and their
derivatives with respect to the scale size of an universe $a$ (prime denotes
$\frac{d}{da}$)%
\begin{equation}
H^{2}-\frac{2\omega }{3}\,F^{2}+2\,H\,F+\frac{1}{a^{2}}-\frac{2\omega }{3}%
\,m^{2}=\left( \frac{4\omega }{3}\right) \frac{\rho _{M}}{\phi ^{2}}
\label{HF1**}
\end{equation}%
\begin{equation}
H^{2}+\left( \frac{2\omega }{3}+\frac{4}{3}\right) \,F^{2}+\frac{4}{3}\,H\,F+%
\frac{2a}{3}\,\left( H\,\acute{H}+H\,\acute{F}\right) +\frac{1}{3a^{2}}-%
\frac{2\omega }{3}\,m^{2}=\left( \frac{-4\omega }{3}\right) \frac{p_{M}}{%
\phi ^{2}}  \label{hf2**}
\end{equation}%
\begin{equation}
H^{2}-\frac{\omega }{3}\,F^{2}-\omega \,H\,F+a\left( \frac{H\,\acute{H}}{2}-%
\frac{\omega }{3}\,H\acute{F}\right) +\frac{1}{2a^{2}}-\frac{\omega }{3}%
\,m^{2}=0.  \label{hf3}
\end{equation}

Solving the field equations (\ref{des}-\ref{fi}) for the closed universe
stable-vacuum solution, we get $\phi =\phi _{p}\,e^{F_{p}t}$ and $a\left(
t\right) =a_{\ast }\approx 1/\sqrt{\omega }m$ where $\ F_{P}\approx 0.7m$ is
the primordial fractional rate of change of $\phi $, $a_{\ast }$ is the
constant size of this static universe, $\phi _{p}$ is the constant value of
the field $\phi $ as $t\rightarrow 0$ and $m$ is the mass of the scalar
field $\phi $. The point to note here is that the closed universe ($k=1$)
vacuum solution becomes important for the primordial universe since
homogeneity of the universe only makes sense if a closed universe undergoes
big-bang. Under these closed universe vacuum solutions, we have shown \cite%
{12} how the presence of radiation changes the behavior of the universe
compared to these stable solutions. Putting the scalar field vacuum solution
$\phi \sim \,e^{F_{p}t}$ in (\ref{fi}) and imposing initial conditions as
the big-bang time and the corresponding time dependence of the scale size of
the universe, we have found the solution for the scale size of the universe
in the form of
\begin{equation}
a^{2}\left( t\right) =a_{\ast }^{2}\left[ 1-(1+c)e^{-2F_{p}t}+ce^{2H_{p}t}%
\right]  \label{sol a}
\end{equation}
where $c$ is an integration constant and $H_{p}$ is the primordial Hubble
parameter with $H_{p}=\omega F_{p}\approx $ $0.7\omega m$ for $\omega \gg 1$%
. This solution provides importance for the following reasons:

(1) It is a natural solution. Namely, it does not need any \textquotedblleft
\textit{special\textquotedblright } equation of state for the matter. It is
just deduced from the theory by putting the stable-empty universe solution $%
\phi \sim \,e^{F_{p}t}$ into the equation (\ref{fi}) \cite{12}.

(2) If one examines this inflationary solution concerning as $t\rightarrow 0$
and as $\ t\gtrsim 0$, it is seen that (\ref{sol a}) is both consistent with
$a\left( t\right) \sim \sqrt{t}$ as $t\rightarrow 0$ and also with
primordial rapid inflation described by $a\left( t\right) \sim e^{H_{P}t}$
for $\omega $ $\gg 1$.

(3) We have also checked that for $\omega $ $\gg 1$ and in the limit as $%
t\rightarrow 0$, if one substitutes $\phi \sim e^{F_{P}t}$ and $a\sim \sqrt{t%
}$ into (\ref{des}-\ref{fi}) then the equation of state $p=1/3\rho $ is
satisfied automatically as expected in the radiation dominated epoch of the
standard Einstein cosmology.

In the light of this encouraging result obtained by using the instability
caused by the nonvacuum in the closed stable vacuum solution in explaining
the rapid primordial inflation, we will show that a linearized non-vacuum
solution about the flat stable vacuum solution can also be \ powerful in
explaining the slow late time expansion. Since the universe becomes
(approximately) flat in late times, we ignore the curvature parameter $%
k/a^{2}$ as $a\left( t\right) $ increases with the expansion of the
universe. Under these considerations, in analogy with the assumption we use
in explaining rapid primordial inflation, we first propose $a=e^{H_{\infty
}t}$ and $\phi =e^{F_{\infty }t}$ and put into (\ref{des}-\ref{fi}) and
search for a zeroth order stable vacuum (empty except the $\phi $ field)
solution. $H_{\infty },$ $F_{\infty }$ are the constants to be determined
named as the late time Hubble parameter and the fractional rate of change of
$\phi $ in the late time regime respectively. We have the following coupled
equations for $H_{\infty }$ and $F_{\infty }$;%
\begin{equation}
H_{\infty }^{2}-\frac{2}{3}\,\omega \,F_{\infty }^{2}+2\,H_{\infty
}F_{\infty }-\frac{2\omega }{3}\,m^{2}=0  \label{bta1}
\end{equation}%
\begin{equation}
H_{\infty }^{2}+\left( \frac{2}{3}\omega +\frac{4}{3}\right) \,F_{\infty
}^{2}+\frac{4}{3}\,H_{\infty }F_{\infty }-\frac{2\omega }{3}\,m^{2}=0
\label{beta}
\end{equation}%
\begin{equation}
H_{\infty }^{2}-\frac{\omega }{3}\,F_{\infty }^{2}-\omega \,H_{\infty
}F_{\infty }-\frac{\omega }{3}\,m^{2}=0  \label{beta3}
\end{equation}%
\noindent\ and get the solution;%
\begin{equation}
H_{\infty }=2\,\left( \omega +1\right) \left( \frac{\omega }{6\omega
^{2}+17\omega +12}\right) ^{1/2}\,m\approx 0.8\sqrt{\omega }\,m
\label{Halfa}
\end{equation}
\begin{equation}
F_{\infty }=\left( \frac{\omega }{6\omega ^{2}+17\omega +12}\right)
^{1/2}\,m\approx \frac{0.4}{\sqrt{\omega }}\,m  \label{beta5}
\end{equation}%
where the approximations are again for $\omega \gg 1$. Thus, in our one
hand, we have had an exact zeroth order stable-vacuum solution as,%
\begin{equation}
a=e^{H_{\infty }t}  \label{vacuum a}
\end{equation}%
\begin{equation}
\phi =\phi _{\infty }e^{F_{\infty }t}  \label{vaccum fay}
\end{equation}

\noindent where $H_{\infty }\approx 0.8\sqrt{\omega }m$, $F_{\infty }\approx
\frac{0.4}{\sqrt{\omega }}m$, $\phi _{\infty }$ is a constant.

Then, after finding such a zeroth order exact stable solution, the question
that stimulates us, similar to the primordial regime analysis, is that how
the presence of matter affects this flat stable vacuum solution (\ref{vacuum
a}, \ref{vaccum fay}). To understand such a perturbation phenomenon, we
impose the following linearized first order non-vacuum solution for $H\equiv
\dot{a}/a$ and $F\equiv \dot{\phi}/\phi $ which includes first order
perturbation functions of $h(a)$ and $f(a)$ in addition to the constant
terms $H_{\infty }$ and $F_{\infty }$ which appear in the flat stable vacuum
solution (\ref{vacuum a}, \ref{vaccum fay}) respectively.%
\begin{equation}
H=H_{\infty }+h(a)  \label{linH*}
\end{equation}%
\begin{equation}
F=F_{\infty }+f(a).  \label{lin F}
\end{equation}%
Since solving the field equations (\ref{des}-\ref{fi}) exactly for $a(t)$
and $\phi (t)$ under the condition $p=0$ is hard enough, we put our imposed
solution (\ref{linH*}, \ref{lin F} ) into the modified field equations (\ref%
{hf2**}-\ref{hf3}) for $p=0$ and neglect higher terms in $h(a)$, $f(a)$ then
we get $h(a)$ and $f(a)$ for all $\omega $ in the form of,%
\begin{equation}
h(a)=C_{1}H_{0}\left( \frac{a_{0}}{a}\right) ^{\frac{3\omega +4}{\omega +1}%
}-\left( \frac{1}{H_{\infty }a_{0}^{2}}\right) \frac{(\omega +1)(\omega +3)}{%
(\omega +2)(2\omega +3)}\left( \frac{a_{o}}{a}\right) ^{2}  \label{h(a)}
\end{equation}%
\begin{equation}
f(a)=C_{2}H_{0}\left( \frac{a_{0}}{a}\right) ^{\frac{3\omega +4}{\omega +1}%
}+\left( \frac{3}{2H_{\infty }a_{0}^{2}}\right) \frac{(\omega +1)}{(\omega
+2)(2\omega +3)}\left( \frac{a_{o}}{a}\right) ^{2}  \label{f(a)}
\end{equation}%
where $a_{0}$ is present size of the universe and $H_{0}$ is the present
Hubble parameter. $C_{1}$ and $C_{2}$ are, on the other hand, dimensionless
integration constants. Since letting $\omega \rightarrow \infty $ has a
special meaning in the sense that the Brans-Dicke scalar tensor theory
matches with standard Einstein theory under such limit, we display the
linearized solution (\ref{linH*}, \ref{lin F}) in the following form as $%
\omega \rightarrow \infty $,
\begin{equation}
H=H_{\infty }+C_{1}H_{0}\left( \frac{a_{0}}{a}\right) ^{3+\frac{1}{\omega }}-%
\frac{1}{2}\left( \frac{1}{H_{\infty }a_{0}^{2}}\right) \left( \frac{a_{o}}{a%
}\right) ^{2}  \label{exactH*}
\end{equation}%
\begin{equation}
F=F_{\infty }+C_{2}H_{0}\left( \frac{a_{0}}{a}\right) ^{3+\frac{1}{\omega }}.
\label{exact F}
\end{equation}%
\ Hence, putting the solution (\ref{exactH*}) in the standard Friedmann
equation,%
\begin{equation}
\left( \frac{H}{H_{0}}\right) ^{2}=\Omega _{\Lambda }+\Omega _{R}\left(
\frac{a_{0}}{a}\right) ^{2}+\Omega _{M}\left( \frac{a_{0}}{a}\right) ^{3}
\label{fridmann}
\end{equation}%
which is used for fitting Hubble parameter to the measured density
parameters of universe in such a way that $\Omega _{\Lambda }+$ $\Omega
_{R}+ $ $\Omega _{M}=1$ and using the present observational results on
density parameters $\Omega _{\Lambda }\simeq 0.75$, $\Omega _{M}\simeq 0.25$%
, $\Omega _{R}\simeq 0$ \cite{knop}, we get $C_{1}\simeq 0.15$ and $%
H_{\infty }=0.86H_{0}$ so that $F_{\infty }\approx H_{\infty }/2\omega
\approx \left( 0.43/\omega \right) H_{0}$ which provides $\left\vert
C_{2}\right\vert \ll 0.43/\omega $. Namely, the first term in a linearized
solution (\ref{exact F}) is much greater than the second term. The curvature
density parameter $\Omega _{R}$, on the other hand, is found to be in
accordance with the recent measurements since the term $(1/H_{0}a_{0})^{2}%
\approx $ $\Omega _{R}$ $\simeq 0$ \cite{knop}.

To compare (\ref{HF1**}), with standard Friedmann-Lamaitre cosmology, we put
the linearized solution (\ref{exactH*}, \ref{exact F}) into this equation
and transfer all terms except for $H^{2}=\left( \dot{a}/a\right) ^{2}$ to
the right hand side. Neglecting the $1/a^{2}$ term%
\begin{equation}
H^{2}=\frac{4\omega }{3\phi ^{2}}(\rho _{\Lambda }+\rho _{M}+\rho _{D}).
\label{rho}
\end{equation}%
Noting that, in the late time regime, $\phi \sim a^{1/2\omega }$ is
approximately constant as $a$ changes we identify the terms which do not
explicitly depend on $a$ with $\rho _{\Lambda }$ and terms which depend on $%
a $ as $a^{-3}$ with the dark matter energy density $\rho _{D}$ so that
\begin{equation}
\rho _{D}=(C_{2}F_{\infty }H_{0}\phi ^{2}-\frac{3C_{2}}{2\omega }H_{\infty
}H_{0}\phi ^{2}-\frac{3C_{1}}{2\omega }H_{0}F_{\infty }\phi ^{2})(\frac{a_{0}%
}{a})^{3}  \label{rodark}
\end{equation}%
\begin{equation}
\rho _{\Lambda }=\frac{1}{2}F_{\infty }^{2}\phi ^{2}-\frac{3}{2\omega }%
H_{\infty }F_{\infty }\phi ^{2}+\frac{1}{2}m^{2}\phi ^{2}.  \label{rogama}
\end{equation}

Using the recent observational results on density parameters of the universe
$(\Omega _{\Lambda }$ $\equiv \rho _{\Lambda }/\rho _{0}\simeq 0.75,$ $%
\Omega _{D}\equiv \rho _{D}/\rho _{0}\simeq 0.23)$ where $\rho _{0}$ is the
present measured energy density of the universe and the relations $F_{\infty
}\approx H_{\infty }/2\omega \approx \left( 0.43/\omega \right) H_{0}$ and $%
H_{\infty }\approx 0.8\sqrt{\omega }m$ as $\omega \rightarrow \infty $, we
fit (\ref{rodark}, \ref{rogama}) to the ratio $\Omega _{\Lambda }/\Omega
_{D} $ $\simeq 75/23$ and determine the $\left\vert C_{2}\right\vert $
integration constant to be $\left\vert C_{2}\right\vert \approx 0.20$ which
is inconsistent with the requirement $\left\vert C_{2}\right\vert \ll
0.43/\omega $ imposed by the theory.

In conclusion, the first remarkable feature of this work that a linearized
non-vacuum solution (\ref{exactH*}) about the stable cosmological vacuum
solution (\ref{Halfa}) with flat $(k=0)$ space-like section is capable of
explaining how the Hubble parameter $H\equiv \dot{a}/a$ evolves with the
scale size of the universe $a(t)$.

The second remarkable feature of this theory is that by fitting the
linearized solutions (\ref{exactH*}, \ref{exact F}) of the theory to the
recent observations \cite{knop}, the late-time Hubble parameter $H_{\infty
}=0.86H_{0}$ and the fractional rate of change of $\phi $ in the late time
regime $F_{\infty }=$ $\left( 0.43/\omega \right) H_{0}$ are successfully
predicted in terms of today's observational measured vale of Hubble
parameter $H_{0}$.

Another important prediction we note from this theory is that for a fixed $%
H_{0}$, since $F_{\infty }\approx \left( 0.43/\omega \right) H_{0}$, $%
F_{\infty }$ may not attain a large value because of its inverse dependence
on $\omega $ which is measured to be, according to the recent observational
data, as $\omega >10^{4}\gg 1$ \cite{omega, omega1}. This is the reason why $%
F_{\infty }$ can not let the scalar field $\phi =\phi _{\infty }e^{F_{\infty
}t}$ to blow up rapidly so that $\rho _{\Lambda }$ (\ref{rogama}), the
energy density of the universe due to the cosmological constant, can grow
slowly and reasonably. Hence we strictly agree on that this theory is
successful in explaining the dark energy though the Brans-Dicke scalar field
$\phi $ can not account for dark matter.

The last remarkable feature of this theory is that it enables us to estimate
some dimensionful parameters displayed in the theory. Using the relation $%
H_{\infty }\approx 0.86H_{0}\approx 0.8\sqrt{\omega }m$ and the restriction
on $\omega $, we may estimate $m$ for a fixed $H_{0}$ as%
\begin{equation}
m\lessapprox 10^{-2}H_{0}  \label{mass}
\end{equation}%
\noindent where the present value of Hubble constant $H_{0}=720\pm 8$ km s$%
^{-1}$ Mpc $^{-1}$ \cite{freedman}. Using appropriate conversion relation in
relativistic units, present Hubble parameter is found to be $H_{0}$ $\approx
10^{-26}$m$^{-1}\approx 2\times 10^{-42}$ GeV.

By using the relation $G_{eff}^{-1}=(2\pi /\omega )\phi ^{2}$, the time
variation in Newtonian gravitation constant can be written as $\mid $ $\dot{G%
}/G\mid =2F=2\left( \dot{\phi}/\phi \right) $ so that we can estimate the
time variation in Newtonian constant $G$, $F_{P}$, $F_{\infty }$ for the
primordial and the late time epochs considered in this theory as

\begin{equation}
|\frac{\dot{G}}{G}|_{P}=2F_{P}\approx (1.5/\sqrt{\omega })H_{0}
\label{gprimor}
\end{equation}
\begin{equation}
|\frac{\dot{G}}{G}|_{\infty }=2F_{\infty }\approx \frac{0.86}{\omega }\,H_{0}
\label{Glate}
\end{equation}%
\begin{equation}
F_{P}\lessapprox \,7\times 10^{-3}\,H_{0}  \label{fp}
\end{equation}%
\begin{equation}
F_{\infty }<43\times 10^{-6}H_{0}.  \label{fsonsuz}
\end{equation}
\noindent The Hubble parameters, on the other hand, are given by
\begin{equation}
H_{P}\approx 0.7m\omega >70\,H_{0}  \label{Hprim}
\end{equation}
\begin{equation}
H_{\infty }\approx 0.8\sqrt{\omega }m\approx \allowbreak 0.86H_{0}.
\label{Hsonsuz}
\end{equation}

We would like to thank the anonymous referee for thoughtful comments and
valuable suggestions which provided to improve and to clarify the paper.
This work is supported by Bogazici University Research Fund, Project no:
04B302D.

\end{document}